\documentclass[aps, prb, twocolumn, longbibliography, preprintnumbers, floatfix]{revtex4-2}
\usepackage{amssymb,amsfonts,amsmath,dsfont}
\usepackage{graphicx} 
\usepackage{textcomp}
\usepackage{gensymb}
\usepackage{pifont}
\usepackage{bm} 
\usepackage{color}
\usepackage{accents}
\usepackage{mathtools}
\usepackage{hyperref}
\usepackage{booktabs}
\usepackage{nicefrac}
\usepackage{url}
\hypersetup{colorlinks=true, citecolor=blue, urlcolor=blue, linkcolor=blue}
\usepackage{multirow}
\usepackage{svg}

\usepackage{comment}

\begin{document}
\title{Tunable topological magnon-polaron states and intrinsic anomalous Hall phenomena in two-dimensional ferromagnetic insulators}

\author{Jostein N. Kløgetvedt}
\author{Alireza Qaiumzadeh}
\affiliation{Center for Quantum Spintronics, Department of Physics, Norwegian University of Science and Technology, NO-7491 Trondheim, Norway}

\date{\today}

\begin{abstract}
We study magnon-polaron hybrid states, mediated by Dzyaloshinskii-Moriya and magnetoelastic interactions, in a two-dimensional ferromagnetic insulator. 
The magnetic system consists of both in-plane and flexural acoustic and optical phonon bands, as well as acoustic and optical magnon bands. Through manipulation of the ground-state magnetization direction using a magnetic field, we demonstrate the tunability of Chern numbers and (spin) Berry curvatures of magnon-polaron hybrid bands.  
This adjustment subsequently modifies two intrinsic anomalous Hall responses of the system, namely, the intrinsic thermal Hall and intrinsic spin Nernst signals.
Notably, we find that by changing the magnetic field direction in particular directions, it is possible to completely suppress the thermal Hall signal while maintaining a finite spin Nernst signal.
Our finding reveals the intricate interplay between topology and magnetic ordering, offering compelling avenues for on-demand control over emergent nontrivial topological states and quantum transport phenomena in condensed matter systems by potential applications in both classical and quantum information technology.
\end{abstract}

\maketitle

\section{Introduction}
Magnon-polarons are emergent quasiparticles arising from hybrid states between magnons, the quanta of collective spin excitations, and phonons, the quanta of lattice vibrations \cite{10.1063/5.0047054,Bao2023,PhysRevB.102.144438, physics,Bozhko_2020}. The coherent coupling between magnons and phonons modifies the thermoelectric responses of the material. By studying magnon-polaron quasiparticles, we may gain insight into the interaction strength and nature of the spin-phonon coupling, which in turn can reveal information about the ground state of the system.
On the other hand, these emerging hybrid modes, with typical subnanometer wavelengths, may exhibit coherent spin angular momentum transport over long distances \cite{Zhang2021,Bauer}, and can also manifest nontrivial topological properties \cite{PhysRevLett.117.217205,Bao2023}. 
These features make magnon-polaron hybrid modes promising for the realization of functional hybrid quantum systems \cite{Lachance-Quirion_2019} with potential applications in low-power and high-speed spintronic nanodevices, compact topological devices, hybrid quantum systems and quantum information technology. Therefore, the tunability of coherent coupling between magnons and phonons is an essential prerequisite for their application in modern quantum technology \cite{Bao2023}.

Recently discovered two-dimensional (2D) ferromagnetic (FM) and antiferromagnetic (AFM) materials \cite{Burch,Rodin_2020, doi:10.1021/acsnano.1c06864,doi:10.1021/acsaelm.2c00419, Gibertini_2019} are ideal platforms and testbeds for investigation of these emerging quasiparticles. 
Theoretical and experimental studies in 2D FM \cite{PhysRevLett.117.217205,PhysRevB.106.214424,PhysRevB.107.214452,PhysRevB.104.045139,PhysRevLett.122.107201,PhysRevLett.123.167202,PhysRevLett.123.237207,PhysRevB.104.064305,PhysRevB.107.184434,PhysRevLett.129.067202,PhysRevB.101.064424,PhysRevB.106.125103,PhysRevLett.127.247202} and AFM \cite{PhysRevB.104.L180402,PhysRevLett.127.097401,PhysRevB.99.174435,PhysRevB.104.134437,PhysRevB.105.L100402,PhysRevLett.124.147204,doi:10.1021/acs.nanolett.3c00351,doi:10.7566/JPSJ.88.081003,PhysRevB.107.L060404,to2023giant, Thermal_Hall_Effect_Spin_Nernst_Effect_Spin_Density,Bao2023,PhysRevB.108.L140402} systems have shown that the coherent interaction between magnons and phonons may generate nontrivial topological bands from trivial magnon and phonon bands, or strengthen the already existing topological bands, and consequently lead to various anomalous Hall effects, mediated by emerging magnon-polaron quasiparticles. These phenomena originate from finite Berry curvatures, generated in level repulsion hotspots where magnon and phonon branches intersect in the presence of coherent interactions \cite{PhysRevLett.117.217205}. 

There are various mechanisms for coherent magnon-phonon (m-ph) coupling in collinear and noncollinear magnetic insulators, such as  dipolar interactions \cite{PhysRevLett.117.217205}, spin-orbit interactions, including Dzyaloshinskii-Moriya (DM) interactions and magnetoelastic couplings \cite{10.1063/5.0047054,PhysRev.110.836,10.1063/1.1735909,PhysRevLett.123.167202,PhysRevB.89.184413}, Heisenberg exchange interactions \cite{PhysRevB.99.174435}, and a magnetic field gradient \cite{Nunez}.

The fingerprint of nontrivial topology and finite Berry curvatures in the ground-state and/or excitation properties of a system can be intercepted in Hall-type quantum transport phenomena, such as anomalous thermal Hall and spin Nernst effects.  
The tuning of Chern numbers, Berry curvatures, and quantum transports is highly demanded for application in emerging quantum technology \cite{Bao2023}.

Recently, it was shown theoretically that an effective magnetic field or magnetic anisotropy can tune Chern numbers and the thermal Hall effect of magnon-polarons, generated via magnetoelastic interactions, in FM systems by changing the number of band-crossing lines \cite{PhysRevB.106.125103, PhysRevLett.123.237207}. 
In Ref. \cite{PhysRevB.101.125111}, the authors studied how varying the direction of a magnetic field may change the topological structure and thermal Hall response of magnon-acoustic phonon hybrid modes in a simple square lattice with the Kittel-type magnetoelastic hybridization mechanism \cite{PhysRev.110.836,RevModPhys.21.541}. 

In this paper, we demonstrate the influence of tuning the direction of an external magnetic field on the topological properties of magnon-polaron hybrid bands in a FM system with and without broken spatial inversion symmetry. We examine the intricate hybridization of magnons with both in-plane (IP) and out-of-plane (OOP) or flexural modes of acoustic and optical phonons in the presence of two types of spin-orbit mediated hybridization mechanisms: generalized DM interactions and the Kittel-type magnetoelastic interactions.
We explore the fingerprint of these topological properties through their influence on thermal Hall and spin Nernst effects.

Focusing only on the magnon-polaron contribution, we show that by tuning the ground-state magnetization direction, we can turn off the thermal Hall signal while the spin Nernst signal remains finite. This behavior of magnon-polaron hybrid modes in our FM system, resembles pure magnonic counterpart effects in collinear AFM honeycomb lattices. In such systems, the thermal Hall effect is forbidden by symmetry, whereas the spin Nernst Hall effect remains finite \cite{PhysRevLett.117.217203, PhysRevLett.117.217202}. 

The rest of the paper is organized as follows. We first introduce our effective model Hamiltonian with different m-ph coupling mechanisms in Sec. \ref{Model}. In Sec. \ref{m-ph}, we present the band structure of magnon-polaron modes in the presence of various m-ph coupling mechanisms. Chern numbers of different magnon-polaron hybrid bands are computed in Sec. \ref{Chern}. Thermal Hall and spin Nernst conductivities are computed in Sec. \ref{transport}. We then summarize our findings in Sec. \ref{summary} and discuss their implications.

\begin{figure}
    \centering
    \includegraphics[width=1\linewidth]{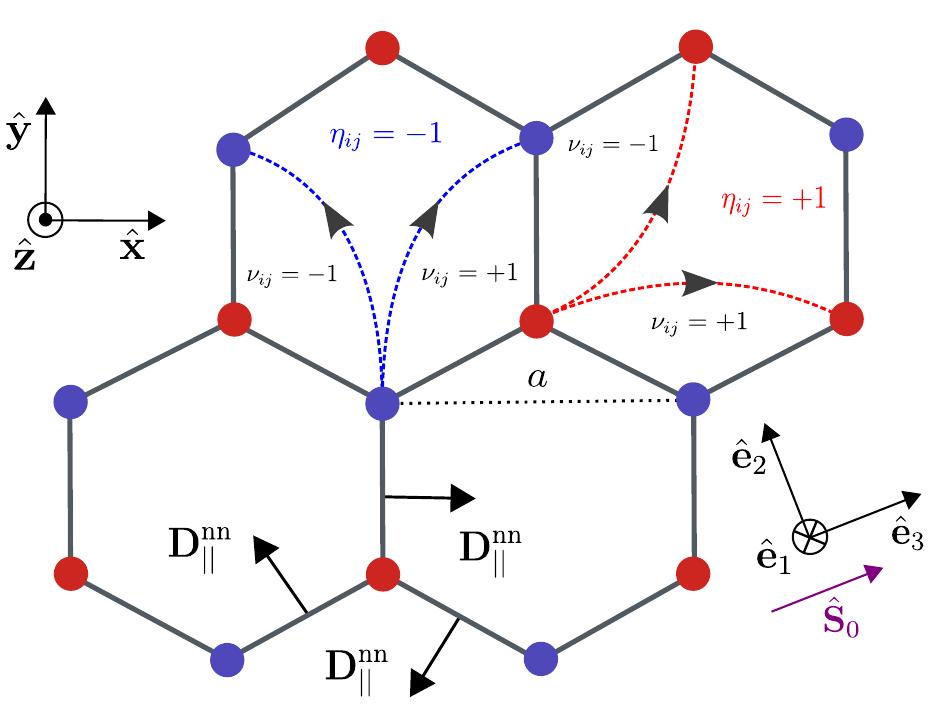}
    \caption{Schematic representation of a honeycomb lattice. Red and blue vertices represent $A$ and $B$ sublattices, respectively. $\nu_{ij}=+ (-)$ for clockwise (counterclockwise) nnn hopping while $\eta_{ij}=+ (-)$ for nnn hopping between A (B) sublattices. $\bm{D}_{||}^{\text{nn}} = D_{xy}^{\text{nn}} ( \hat{\bm{z}} \times \hat{\bm{R}}_{ij})$ is the IP nn DM vector, $\hat{\bm{R}}_{ij}$ is the nn vector, and $a$ is the lattice constant.
     $\hat{\bm{x}}-\hat{\bm{y}}-\hat{\bm{z}}$ is the laboratory coordinate frame while
     $\hat{\bm{e}}_1-\hat{\bm{e}}_2-\hat{\bm{e}}_3$ is the rotated coordinate frame. $\hat{\bm{S}_0}$ is the magnetic ground-state direction.}
    \label{fig: honeycomb_figure}
\end{figure}

\section{Model}\label{Model}
We consider an FM spin Hamiltonian $\mathcal{H}=\mathcal{H}_{\mathrm{m}}+\mathcal{H}_{\mathrm{ph}}+\mathcal{H}_{\mathrm{m-ph}}$, consisting of the Hamiltonian of localized spins $\mathcal{H}_{\mathrm{m}}$, phonons $\mathcal{H}_{\mathrm{ph}}$, and the m-ph interaction $\mathcal{H}_{\mathrm{m-ph}}$. Without loss of generality, we consider a honeycomb lattice structure.

\subsection{Spin Hamiltonian}
The spin Hamiltonian reads \cite{Biquadratic},
\begin{align}
\label{eqn: magnon_general_spin_in_plane_Hamiltonian}
    \mathcal{H}_{\mathrm{m}} =& - J \sum_{\langle i,j \rangle} \bm{S}_i \cdot \bm{S}_j - \Lambda \sum_{\langle i,j \rangle} (\bm{S}_i \cdot \bm{S}_j)^2 - K_z\sum_i S_{iz}^2 \nonumber\\ & - \sum_{i,j} \bm{D}_{ij} \cdot [\bm{S}_i \times \bm{S}_j]- \sum_i \bm{S}_{i} \cdot \bm{h},
\end{align}
where $\bm{S}_i$ is the vector spin operator at the site $i$ with amplitude $|\bm{S}_i|=S$, $J>0$ and $\Lambda>0$ parametrize the isotropic bilinear FM Heisenberg and biquadratic exchange interactions, respectively, between nearest neighbor (nn) sites, $K_z>0$ parametrizes the single-ion easy-axis magnetic anisotropy, $\bm{D}_{ij}$ denotes the DM vector between both nn and next-nearest-neighbor (nnn) sites, and $\bm{h}=g \mu_B \bm{B}$ is the Zeeman field, with $g$ being the Land{\'e} g-factor, $\mu_B$ being the Bohr magneton, and $\bm{B}$ being the external magnetic field. 
In general, we define the following generalized nn and nnn DM vectors \cite{PhysRevB.103.174410,PhysRevLett.123.167202}:
\begin{subequations}
\begin{align}
\label{nnnDMI}
    &\bm{D}^{\text{nnn}}_{ij} = \nu_{ij}D_z^{\text{nnn}} \hat{\bm{z}}-\eta_{ij} D_{xy}^{\text{nnn}} \hat{\bm{R}}_{ij} , \\
    \label{nnDMI}
    &\bm{D}^{\text{nn}}_{ij} =D_z^{\text{nn}} \hat{\bm{z}}+ D_{xy}^{\text{nn}} (\hat{\bm{z}} \times \hat{\bm{R}}_{ij}),
\end{align}
\end{subequations}
where $\nu_{ij}=-\nu_{ji}=\pm 1$ depending on the hopping orientation from site $i$ to nnn site $j$, $\eta_{ij} = +1 (-1)$ for bonds between $A (B)$ sublattices, see Fig \ref{fig: honeycomb_figure}, and $\hat{\bm{R}}_{ij}$ represents the unit vector connecting the lattice sites $i$ and $j$. In a freestanding 2D honeycomb  lattice, only the OOP nnn DM vector can be finite \cite{PhysRevB.100.060410}, while the other DM vectors might be finite when the spatial inversion and/or mirror symmetries are broken \cite{PhysRevMaterials.4.094004}. Realistic materials can exhibit the breaking of these symmetries through growth on different substrates or by applying a gate voltage.
Using the Holstein-Primakoff transformation \cite{HP}, the noninteracting magnon Hamiltonian in the second quantized representation, for an arbitrary direction of the ground-state magnetization, see Appendix \ref{AppA}, reads, 
\begin{align}
    \mathcal{H}_{\mathrm{m}} = \sum_{\bm{q},\sigma} E_{\bm{q}\sigma} a_{\bm{q},\sigma}^\dagger a_{\bm{q},\sigma} ,   
\end{align}
where $a_{\bm{q},\sigma} (a^{\dagger}_{\bm{q},\sigma})$ is the bosonic annihilation (creation) operator for acoustic, $\sigma=-$, and optical, $\sigma=+$, magnon modes with eigenenergy $E_{\bm{q}\sigma}$ \cite{Jostein}. 
The dispersion of a freestanding FM honeycomb lattice, in the presence of an OOP magnetic field, is reduced to $E_{\bm{q}\sigma}=S\left(\mathcal{Z} \tilde{J} +\sigma \sqrt{[D_z^{\text{nnn}}(\bm{q})]^2+\tilde{J}^2|f(\bm{q})|^2}\right)+\big(h_z+(2S-1)K_z\big)$, 
where $f(\bm{q})=\sum_{i=1}^\mathcal{Z} e^{i \bm{q}\cdot\bm{\delta}_i}$ is the lattice structure factor, $\bm{\delta}_i$ is the $i^{\text{th}}$ nn vector,
$\mathcal{Z}=3$ is the coordinate number of honeycomb lattices, $\tilde{J}=J+2\Lambda S^2$ is the effective exchange interaction, and $D_z^{\text{nnn}}(\bm{q}) = -2 D_z^{\text{nnn}} \sum_{i=1}^{\mathcal{Z}} \sin{(\bm{q} \cdot {\bm{\tau}}_i)}$, with ${\bm{\tau}}_1 = a\hat{\bm{x}}$, ${\bm{\tau}}_2 = a(-\hat{\bm{x}} + \sqrt{3} \hat{\bm{y}})/2$ and ${\bm{\tau}_3} = -a(\hat{\bm{x}} + \sqrt{3} \hat{\bm{y}})/2$, where $a$ is the lattice constant. 

\begin{figure}
  \centering
    \includegraphics[width=1\linewidth]{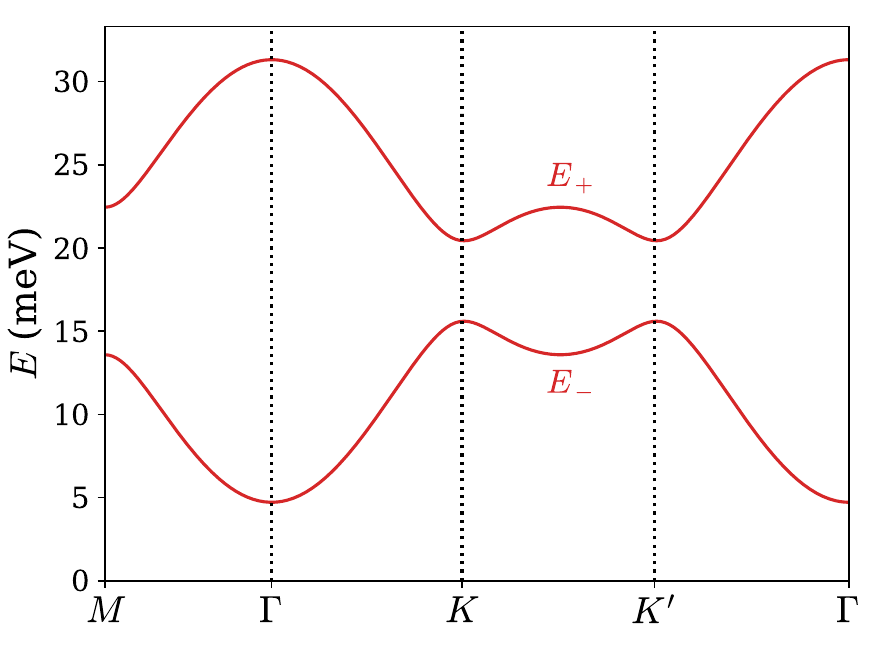}
    \caption{Noninteracting acoustic-like $E_-$ and optical-like $E_+$ magnon bands in an FM honeycomb lattice, in the presence of an OOP magnetic field with amplitude $|\bm{h}|=4.5$ meV. We use the following parameters for CrI$_3$ \cite{Biquadratic,Jostein}: $J = 2.01$ meV, $\Lambda = 0.21$ meV, $K_z = 0.109$ meV, $D_z^{\text{nnn}}=0.31$ meV, and $S=3/2$.}
    \label{fig: magnon_bands}
\end{figure}

Figure \ref{fig: magnon_bands} shows this dispersion relation. There is a topological Dirac-like gap between the acoustic and optical magnon bands at $K$ and $K'$ points in the system with an OOP magnetic ground state due to a finite $D_z^{\text{nnn}}$ \cite{review,PhysRevLett.117.227201}. A similar topological gap has recently been reported in the magnon dispersion of $\text{CrI}_3$ \cite{PhysRevX.8.041028,PhysRevX.11.031047}. The band gap in the acoustic branch of the magnon dispersion at the $\Gamma$ point is due to the OOP magnetic anisotropy and the applied OOP magnetic fields. 

We should emphasize that our linear spin-wave calculation within the harmonic approximation is valid at low temperature, where magnon-magnon interactions are small. At higher temperatures, the number of thermal magnons is increased, and thus nonlinear magnon interactions may play an important role in topological properties of the system \cite{PhysRevLett.127.217202}.

\subsection{Phonon Hamiltonian}
The phonon Hamiltonian reads \cite{Falkovsky1,Falkovsky2},
\begin{align}
    \label{eqn: phonon_Hamiltonian_harmonic}
    \mathcal{H}_{\mathrm{ph}} = \frac{1}{2} \sum_{i, \alpha, \mu} M_{\alpha} \dot{u}_{i\alpha \mu}^2 + \frac{1}{2} \sum_{\substack{i, \alpha, \mu \\ j, \beta, \nu}} u_{i\alpha \mu}\Phi_{\mu \nu}^{\alpha \beta}(\bm{R}_{ji})  u_{j \beta \nu}, 
\end{align}
where $i (j)$ labels the unit cells, $\alpha (\beta)$ labels the ions inside a unit cell, $\mu (\nu)$ denotes spatial coordinates, $M_\alpha$ is the mass of $\alpha$th ion in the unit cell, $\bm{u}_{i\alpha}(t)$ represents the displacement of the ion at time $t$, and $\Phi_{\mu \nu}^{\alpha \beta}$ is the force coefficient between the $i\alpha$th ion in the $\mu$th direction due to a displacement of the $j\beta$th ion in the $\nu$th direction.

\begin{figure}
  \centering
    \includegraphics[width=1\linewidth]{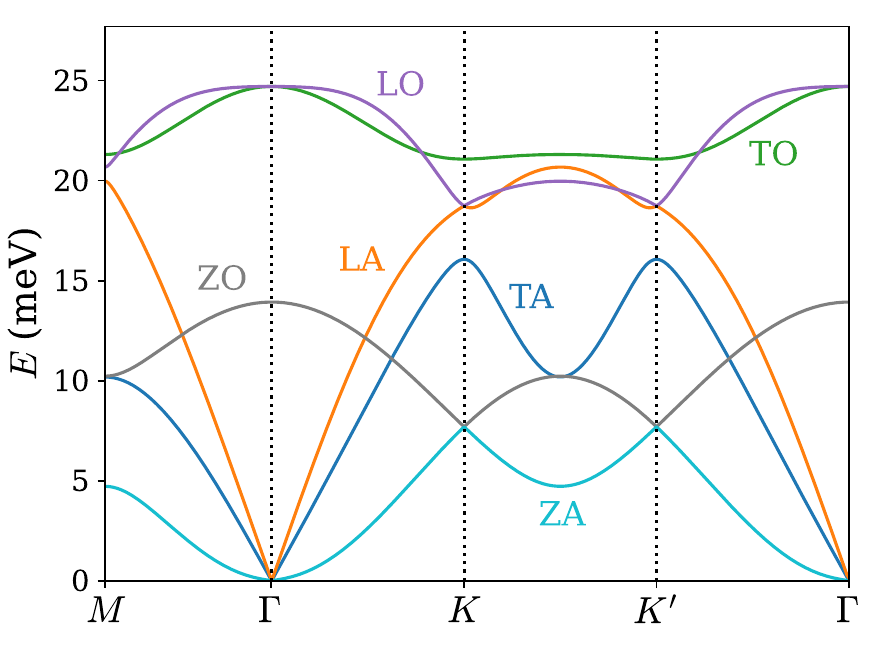}
    \caption{Noninteracting phonon bands in a honeycomb lattice. We use graphene parameters obtained in Refs. \cite{Falkovsky1,Falkovsky2,Jostein}.}
    \label{fig: phonon_bands}
\end{figure}

The second quantized form of the phonon Hamiltonian in the diagonalized basis within the harmonic approximation reads, 
\begin{align}
    \mathcal{H}_{\mathrm{ph}} = \sum_{\bm{q},\lambda} E_{\bm{q}\lambda} \big(c_{\bm{q},\lambda}^\dagger c_{\bm{q},\lambda}+\frac{1}{2}\big),   
\end{align}
where $c_{\bm{q},\lambda} (c^\dagger_{\bm{q},\lambda})$ is the bosonic annihilation (creation) operator for the $\lambda$ phonon mode with the eigenenergy $E_{\bm{q}\lambda}$.
Figure \ref{fig: phonon_bands} shows the phonon dispersion relations of a honeycomb  lattice. This dispersion is consistent with recent theoretical and experimental findings, indicating that graphene sheets contain four distinct types of topological phononic Dirac points and a single phononic nodal ring in their phonon spectrum \cite{TopologicalPhonons,PhysRevLett.131.116602}. There are three acoustic modes: IP longitudinal (LA), IP transverse (TA), and OOP (ZA). Although the two IP acoustic modes have linear dispersions around the $\Gamma$ point, the OOP or flexural acoustic mode in 2D systems has a quadratic low-energy dispersion in the presence of rotation symmetry \cite{PhysRevLett.100.076801,Rodin_2020,PhysRevLett.108.251602}. In addition, there are three optical modes: IP longitudinal (LO), IP transverse (TO), and OOP (ZO).

\subsection{Magnon-phonon interactions}
The  m-ph interactions in our model can in general arise from bilinear and biquadratic exchange interactions, DM interactions, and crystalline magnetic anisotropy. 
We seek coherent hybridization between magnons and phonons that leads to level repulsion between two bosonic modes at degenerate points, rather than scattering phenomena between them. Therefore, we focus solely on the identification of linear-order m-ph interactions. The magnetic ground state is collinear in our model; therefore, up to the lowest order in m-ph coupling amplitude, only IP components of the DM interactions and crystalline magnetic anisotropy may lead to m-ph hybridization. 
Note that the OOP components of DM interactions may also lead to the m-ph hybridization but their contribution is higher order than the in-plane components, see the Appendix \ref{AppB}, and thus we ignore them in this paper.
Therefore, the total m-ph Hamiltonian for an arbitrary direction of the magnetic ground state consists of three terms $\mathcal{H}_{\mathrm{m-ph}}=\mathcal{H}_{\text{D}^{\text{nnn}}_{xy}}+\mathcal{H}_{\text{D}^{\text{nn}}_{xy}}+\mathcal{H}_{\text{me}}$. Look at Appendix  \ref{AppB} for more details.

It is worth mentioning that, as is shown in the following, the different coherent m-ph interaction mechanisms couple spins to the displacement fields, effectively acting as a spin-orbit interaction in the effective Hamiltonian of the magnon-polaron quasiparticles.

(i) The IP nnn-DM interaction, Eq. (\ref{nnnDMI}), leads to an effective m-ph coupling as,
\begin{align}
     \mathcal{H}_{\text{D}^{\text{nnn}}_{xy}} = \sum_{\langle \langle i,j \rangle\rangle} \sum_{\mu,\nu} (u_{i \mu} - u_{j\mu}) F^{\mu\nu}_{ij}  (\bm{S}'_i \times \bm{S}'_j)_{\nu},
\end{align}
with the following coupling matrix for arbitrary spin direction, see Appendix \ref{AppB-1} \cite{Jostein}:
\begin{align}
F^{\mu\nu}_{ij}=\frac{\eta_{ij} D_{xy}^{\text{nnn}}}{|\bm{R}_{ij}|} 
 \sum_{\xi}\left(\delta_{\mu\xi}- \hat{R}_{ij}^{\mu}\hat{R}_{ij}^{\xi} \right) \mathcal{R}_\xi^\nu.
\end{align}
The rotation matrix $\mathcal{R}$ is defined in such a way that spins can be expressed in terms of a new frame coordinates $\{\hat{\bm{e}}_1, \hat{\bm{e}}_2, \hat{\bm{e}}_3\}$, where $\hat{\bm{e}}_3$ aligns with the magnetic ground state, dictated by the magnetic field direction; that is, $(\bm{S}_i \times \bm{S}_j)_{\nu} = [\mathcal{R} (\bm{S}'_i \times \bm{S}'_j)]_{\nu}$ (see Appendix \ref{AppA}). This interaction term can be finite for both in-plane ($\mu=x,y$) and out-of-plane ($\mu=z$) phonon modes.

(ii) The IP nn-DM interaction, Eq. (\ref{nnDMI}), results in the following m-ph coupling,
\begin{align}
    \mathcal{H}_{\text{D}^{\text{nn}}_{xy}} = \sum_{\langle i,j \rangle} \sum_{\mu, \nu} (u_{i\mu}- u_{j\mu}) T_{ij}^{\mu\nu} (\bm{S}'_i \times \bm{S}'_j)_{\nu},
\end{align}
with the nn coupling matrix for arbitrary spin direction being given by \cite{Jostein}
\begin{align}
    T_{ij}^{\mu\nu} = -\frac{D_{xy}^{\text{nn}}}{|\bm{R}_{ij}|}\sum_{\substack{\xi,\gamma}}  \epsilon_{z\gamma \xi} \left(\delta_{\mu\gamma} - \hat{R}_{ij}^{\mu} \hat{R}_{ij}^{\gamma} \right) \mathcal{R}_\xi^\nu,
\end{align}
where $\epsilon_{z\gamma\xi}$ is the Levi-Civita tensor. See Appendix \ref{AppB-2} for details. This interaction term can only be finite for in-plane ($\mu=x,y$) phonon modes.

(iii) The magnetoelastic interaction, the interaction between the spin
and the elastic displacement, arising from the crystalline anisotropy is described by a Kittel-type magnetoelastic energy density at site $i$ as $f^{\text{me}}_i=(b_1/S^2) \sum_\mu e_{\mu\mu} S_{i\mu}^2 + (b_2 / S^2) \sum_{\mu\neq \nu} e_{\mu\nu} S_{i\mu} S_{i\nu}$, where $e_{\mu\nu}=(\partial_{r_\nu} u_\mu+\partial_{r_\mu} u_\nu)/2$ is the strain tensor component and $b_{1}$ and $b_{2}$ are magnetoelastic constants related to the normal strains and shear deformations, respectively \cite{RevModPhys.21.541,PhysRev.110.836}. 
The effective m-ph coupling Hamiltonian in the linear order of magnon amplitude reads \cite{PhysRev.110.836,PhysRevLett.123.237207,PhysRevB.101.125111},
\begin{align}
\mathcal{H}_{\text{me}}=\sum_{\langle i,j \rangle} \sum_{\mu\nu} (u_{i\mu} - u_{j\mu}) K_{ij}^{\mu\nu} S'_{i\nu},
\end{align}
where $K_{ij}$ is the coupling matrix between the $i$ and $j$ sites. For an arbitrary magnetization direction, this matrix is lengthy and is presented in Appendix \ref{AppB-3}. Here, we only consider a magnetization along the $x$, $y$, and $z$ directions, and thus only the shear deformation contributes to the coupling matrix \cite{Jostein},
\begin{align}
    K_{ij} = \frac{\kappa_2}{|\bm{R}_{ij}|^2} \begin{bmatrix}
          R_{ij}^y \Gamma^1_{xy} & R_{ij}^y \Gamma^2_{xy} \\
        R_{ij}^x \Gamma^1_{xy} & R_{ij}^x \Gamma^2_{xy} \\
        R_{ij}^x \Gamma^1_{xz} + R_{ij}^y \Gamma^1_{yz} & R_{ij}^x \Gamma^2_{xz} + R_{ij}^y \Gamma^2_{yz}
    \end{bmatrix},
\end{align}
where $\Gamma^\nu_{\mu\mu'}=\big(\mathcal{R}_\mu^\nu \mathcal{R}_{\mu'}^3 + \mathcal{R}_{\mu}^3 \mathcal{R}_{\mu'}^{\nu}\big)/2$ and $\kappa_{2} = 2 b_2 a^3/S$.

\section{Magnon-polaron bands}\label{m-ph}

\begin{figure*}
    \centering
    \includegraphics[width=1\linewidth]{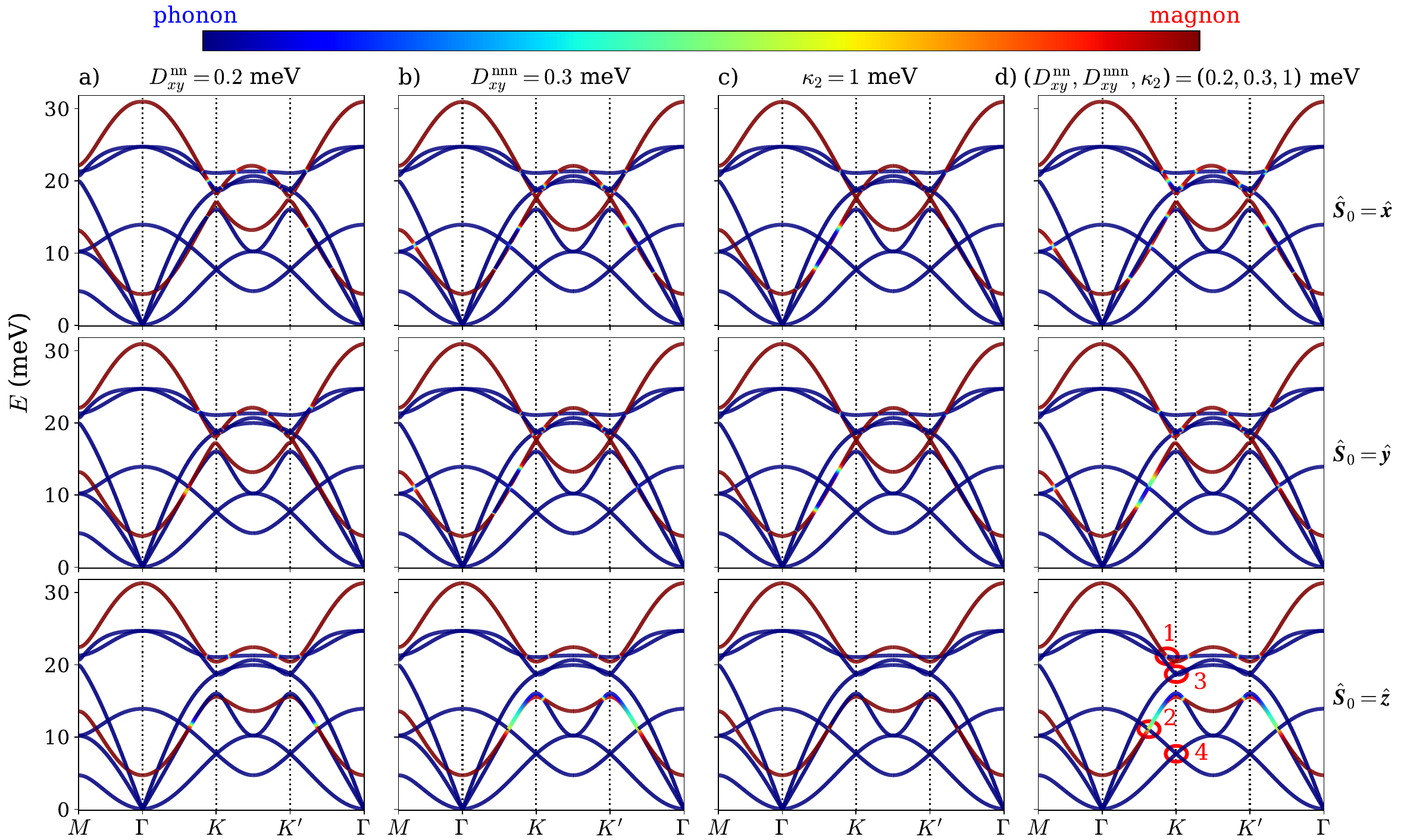}
    \caption{Magnon-polaron bands in an FM insulator with honeycomb lattice structure.
    Plots in each row illustrate the band dispersion with a magnetic ground state $\hat{\bm{S}}_0$ along 
the $\hat{\bm{x}}$, $\hat{\bm{y}}$, and $\hat{\bm{z}}$ directions, controlled by an external magnetic field. 
    In each column, we explore different scenarios with a specific nonzero m-ph coupling parameter: (a) $D_{xy}^{\text{nn}} = 0.2$ meV, (b) $D_{xy}^{\text{nnn}} = 0.3$ meV, and (c) $\kappa_2 = 1$ meV. Additionally, we consider a combined scenario, as depicted in (d), where all the mentioned parameters are present, simultaneously.}
    \label{fig: magnon_polaron_bands}
\end{figure*}
\begin{figure}
  \centering
    \includegraphics[width=1\linewidth]{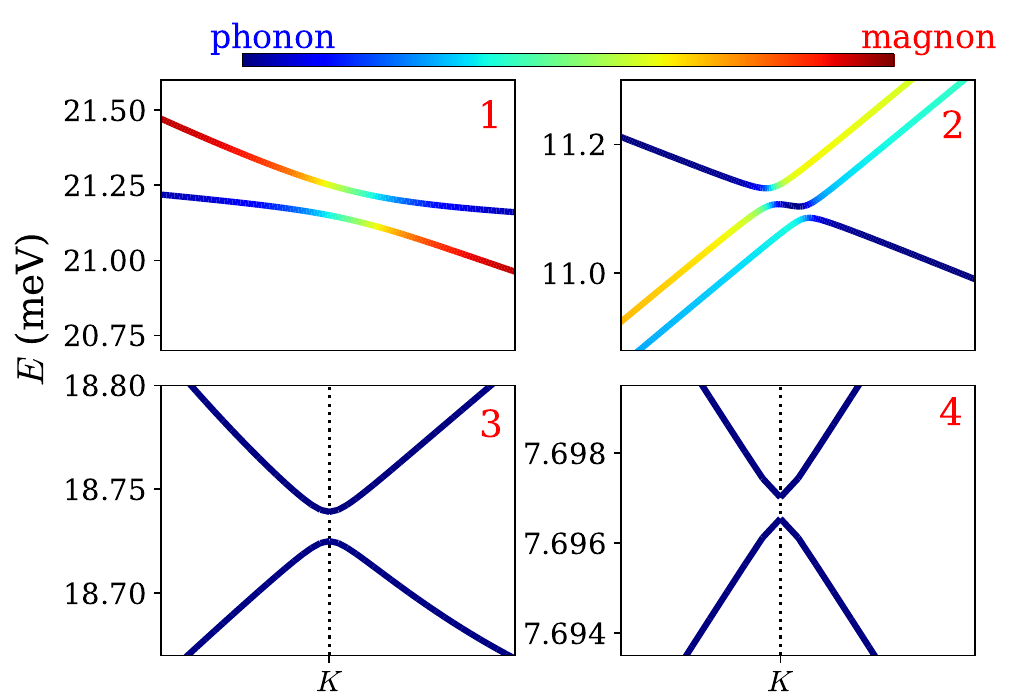}
    \caption{Zoom of the regions around four interaction-induced anticrossing hot spots, as hot spots 1-4; see the bottom right panel in Fig. \ref{fig: magnon_polaron_bands}.  
    The two upper panels show the avoiding crossing between magnon and phonon bands while the two bottom panels show the avoiding crossing between two phonon bands in the presence of coherent m-ph interactions. 
}
    \label{fig: zoom_hybrid_bands}
\end{figure}

We proceed with the diagonalization of the total Hamiltonian, with various m-ph interactions, to determine the eigenstates and eigenenergies associated with the emerging magnon-polaron hybrid modes.
In our model, there are $N_d=8$ magnon-polaron bands with eigenenergies $\mathcal{E}_{k,n}$, where $n=1, ..., N_d$.
In order to explore the impact of different m-ph coupling mechanisms and the orientation of the magnetic ground state on magnon-polaron hybrid bands, we present the band dispersions in Fig. \ref{fig: magnon_polaron_bands}. This figure shows the cases where $D_{xy}^{\text{nn}}$, $D_{xy}^{\text{nnn}}$, or $\kappa_2$ is nonzero, as well as the scenario where all three m-ph mechanisms are present.

We should emphasize that our model Hamiltonian can, in principle, describe a broad range of 2D FM systems. However, in our numerical calculations, we use the DM parameters approximately in the same order as was calculated for CrI$_3$, which is a prototype of 2D FM insulators \cite{Biquadratic, PhysRevB.103.174410}. In the absence of a reported value for the magnetoelastic coupling parameter, $\kappa_2$, in CrI$_3$, and to achieve a response with an amplitude comparable to DM-induced responses, we find that we should assume $\kappa_2$ to be even larger than the DM parameters.

Although a finite $D_{xy}^{\text{nn}}$ does not result in coupling between magnons and OOP phonons for any magnetization direction, hybridization between magnons and OOP phonons may occur when there is a finite IP component of magnetization via $D_{xy}^{\text{nnn}}$. 
On the other hand, both $D_{xy}^{\text{nn}}$ and $D_{xy}^{\text{nnn}}$ induce hybridization between magnons and IP phonons regardless of the magnetization direction.

For a finite magnetoelastic coupling $\kappa_2$, the most profound hybridization occurs with the TA phonon branch when the magnetization lies in the plane, while the coupling to OOP phonon modes is weak in all examined magnetization directions. Furthermore, it is evident that this coupling exhibits stronger hybridization with IP phonon modes near the $\Gamma$ point compared with the DM interactions.

Apart from the avoided level crossings between the magnon and phonon branches, we also observe a tiny gap opening at the intersection between the LO-LA and ZO-ZA phonon branches at the $K$ and $K'$ points in the presence of m-ph interactions; see Fig. \ref{fig: zoom_hybrid_bands}. This gap opens as a consequence of an effective inversion symmetry breaking, $180\degree$ rotation around the $z$ axis, induced by the effective m-ph interactions.
However, these gaps are too small to be visually distinguishable in the presented figures. The presence of such gaps in the phonon spectrum may indicate the existence of chiral phonons, i.e., phonons with circular polarization that have opposite chirality in different valleys ($K$ and $K'$) \cite{Chiral_phonons_at_high_symmetry_points_in_monolayer_hexagonal}. Chiral phonons can possess a finite angular momentum and may exhibit a valley phonon Hall effect. Exploring chiral phonons is beyond the scope of the present study, and we defer this aspect to future investigations.
Similar gap openings have recently been reported in an AFM system \cite{PhysRevB.105.L100402}.


\section{Berry curvature and Chern numbers}\label{Chern}
The Berry curvature is closely related to the topological properties of the energy bands and
plays a crucial role in determining the anomalous transport properties of bosonic systems. Controlling the Berry curvature is important for exploring novel functionalities and potential applications in bosonic topological materials.
The Berry curvature of the $n^{\text{th}}$ band is given by $\Omega_n(\bm{k}) = i \left(\langle{\partial_{k_x} n(\bm{k})} |{\partial_{k_y} n (\bm{k})}\rangle - \langle{\partial_{k_y} n(\bm{k})}|{\partial_{k_x} n(\bm{k})}\rangle \right)$. 
It is more convenient to compute Berry curvatures by transforming the total interacting Hamiltonian into a bosonic Bogoliubov-de Gennes (BdG) form \cite{Jostein, Thermal_Hall_Effect_Spin_Nernst_Effect_Spin_Density}. The bosonic BdG Hamiltonian has two copies of the same eigenstates. In analogy to fermionic systems, we can denote the two sets as particle-like and hole-like states where the states $n=1,\dots,N_d$ represent particle-like bands and the states $n=N_d+1,\dots,2 N_d$ are hole-like bands.
The BdG system is diagonalized with a paraunitary transformation $T_k^\dagger H_k T_k = \mathcal{E}_k = \text{diag}(\mathcal{E}_{k,1},\dots, \mathcal{E}_{k,N_d}, \mathcal{E}_{-k,1}, \dots, \mathcal{E}_{-k,N_d})$ where the matrix $T_k$ satisfies $T_k^\dagger \sigma_3 T_k = \sigma_3$ and $\sigma_3 = \sigma_z \otimes I_{N_d\times N_d}$ \cite{PhysRevResearch.2.013079}. $\sigma_z$ is the $z$ component of the Pauli matrices, and $I_{N_d\times N_d}$ is the unit matrix of dimension $N_d$. The eigenenergies and eigenstates are found numerically using Colpa's method \cite{COLPA1978327}. 
A gauge-independent representation of the Berry curvature reads  \cite{Topological_phases_in_Magnonics_Review}, 
\begin{align}
\label{eqn: Berry_curvature_modified_expr}
        \Omega_n(\bm{k}) = 2 i \hbar^2 \sum_{\substack{m= 1 \\ m \neq n}}^{2 N_d} (\sigma_3)_{nn} (\sigma_3)_{mm} \frac{\langle n_{\bm{k}} |v_x|{m_{\bm{k}}}\rangle \langle{m_{\bm{k}}}|v_y|{n_{\bm{k}}}\rangle}{(\Bar{\mathcal{E}}_{k,n} - \Bar{\mathcal{E}}_{k,m})^2}, 
\end{align}
where $\bm{v}=\hbar^{-1}\partial_{\bm{k}} \mathcal{H}$ is the velocity operator, $\Bar{\mathcal{E}}_{k,n} = (\sigma_3 \mathcal{E}_k)_{nn}$ and $|{n_{\bm{k}}}\rangle$ denotes the eigenstates corresponding to the columns $T_{k,n}$ in the paraunitary matrix. Based on this expression, we deduce that the Berry curvature in the $n^{th}$ band arises from virtual transitions to other bands, $m \neq n$, and the interband coupling is associated with the velocity operator \cite{Cayssol_2021}.

The first Chern number characterizes the topology of the $n^{\text{th}}$ bands and is given by, 
\begin{align}
  C_n=\frac{1}{2\pi}\int_{\text{BZ}}d^2\bm{k} \Omega_n(\bm{k}).  
\end{align}
In the absence of m-ph interactions, phonon bands are topologically trivial in our model. However, it is well known that a finite $D_z^{{\text{nnn}}}$ opens a gap at $K$ points of FM honeycomb lattices and hence topological magnons emerge in two magnon bands by Chern numbers $C=\pm 1$, provided that the magnetization has a nonzero out-of-plane component \cite{Topological_phases_in_Magnonics_Review,2023arXiv230504830Z,Owerre_2016,PhysRevLett.117.227201,review}. 
In contrast, finite $D_{xy}^{\text{nn}}$ and/or $D_{xy}^{\text{nnn}}$ do not open any topological magnon gap in the system, within the linear spin-wave approximation \cite{PhysRevX.11.021061}.
However, upon considering the lowest-order m-ph interactions, the magnon-polaron hybrid states emerge with the possibility of exhibiting nontrivial topological properties. 
The effective Hamiltonian of magnon-polaron states may break combined time-reversal and spin-rotation symmetry around an in-plane axis. 
Topological gaps emerge at anticrossing points of hybrid magnon-polaron bands, or in magnon- (phonon)-like regions of hybrid bands. 
Consequently, the magnon-polaron bands may exhibit a finite Berry curvature, localized around these topological hotspot gaps. These sources of Berry curvature can influence the Chern numbers associated with the bands, contributing to the rich topological aspects of the system.

Table \ref{table: Chern_numbers} presents a comprehensive summary of the Chern numbers associated with distinct bands arising from various m-ph interaction mechanisms and several magnetic ground-state orientations. This table demonstrates that magnon-polaron bands can be either topologically nontrivial ($C_n \neq 0$) or trivial ($C_n = 0$) based on the m-ph coupling mechanism and magnetization direction. 
\begin{table*}
\begin{tabular}{|cc|c|c|c|}
\hline
\multicolumn{2}{|c|}{m-ph coupling mechanism} & $\hat{\bm{S}}_0=\hat{\bm{x}}$ & $\hat{\bm{S}}_0=\hat{\bm{y}}$ & $\hat{\bm{S}}_0=\hat{\bm{z}}$ \\ \hline
\multicolumn{2}{|c|}{no m-ph coupling} & [0, 0, 0, 0, 0, 0, 0, 0] & [0, 0, 0, 0, 0, 0, 0, 0] & [0, 0, 0, 1, 0, 0, 0, -1] \\ \hline
\multicolumn{1}{|c|}{\parbox[t][0.45cm][t]{3mm}{\multirow{4}{*}{\rotatebox[origin=c]{90}{OOP phonons}}}} & $D_{xy}^{\text{nn}}$ & $\times$ & $\times$ & $\times$ \\
\multicolumn{1}{|c|}{} & $D_{xy}^{\text{nnn}}$ & [0, $\times$, $\times$, 0, 0, $\times$, $\times$, 0] & [0, $\times$, $\times$, 0, 0, $\times$, $\times$, 0] & $\times$ \\
\multicolumn{1}{|c|}{} & $\kappa_2$ & $\Omega_n = 0$ & $\Omega_n = 0$ & [0, $\times$, $\times$, -2, 3, $\times$, $\times$, -1]\\
\multicolumn{1}{|c|}{} & $D_{xy}^{\text{nn}} + D_{xy}^{\text{nnn}} + \kappa_2$ & [0, $\times$, $\times$, 0, 0, $\times$, $\times$, 0] & [0, $\times$, $\times$, 0, 0, $\times$, $\times$, 0] & [0, $\times$, $\times$, -2, 3, $\times$, $\times$, -1] \\ \hline
\multicolumn{1}{|c|}{\parbox[t]{3mm}{\multirow{4}{*}{\rotatebox[origin=c]{90}{IP phonons}}}} & $D_{xy}^{\text{nn}}$ & $\Omega_n = 0$ & $\Omega_n = 0$ & [$\times$, 0, 0, -1, $\times$, 3, -3, 1] \\
\multicolumn{1}{|c|}{} & $D_{xy}^{\text{nnn}}$ & [$\times$, 0, -1, 1, $\times$, 3, -2, -1] & [$\times$, 0, 0, 0, $\times$, 0, 0, 0] & [$\times$, 0, 3, -4, $\times$, 2, -2, 1] \\
\multicolumn{1}{|c|}{} & $\kappa_2$ & $\Omega_n = 0$ & $\Omega_n = 0$& $\times$ \\
\multicolumn{1}{|c|}{} & $D_{xy}^{\text{nn}} + D_{xy}^{\text{nnn}} + \kappa_2$ & [$\times$, 0, 1, 0, $\times$, -2, 2, -1] & [$\times$, 0, 0, 0, $\times$, 0, 0, 0]& [$\times$, 0, 0, -1, $\times$, 0, 0, 1] \\ \hline
\end{tabular}
\caption{Chern number, $C_n$, of magnon and phonon bands in the presence of various m-ph coupling mechanisms. The bands are labeled $n=$ [ZA, TA, LA, $E_-$, ZO, LO, TO, $E_+$]. The symbol $\times$ denotes the absence of m-ph coupling. In certain cases, the Berry curvature of the band is zero, $\Omega_n(\bm{k})=0$.} \label{table: Chern_numbers}
\end{table*}

\section{Intrinsic Anomalous Hall responses}\label{transport}
Nontrivial Berry curvature of the energy bands may lead to the emergence of various Hall effects by inducing an anomalous velocity in the corresponding wave packet dynamics.
Measuring various anomalous Hall conductivities in FM systems provides essential information on the underlying state of the system, as well as the nature of emerging quasiparticles in the system \cite{2023arXiv230504830Z,Zink_2022}.
Direct detection of the (spin) Berry curvature and the Chern number is experimentally challenging in bosonic systems. However, the fingerprint of these quantities may be reflected in the anomalous Hall responses to a temperature gradient throughout the system.
In general, the total anomalous Hall responses of a magnetic insulator have three, intrinsic and/or extrinsic, contributions from free magnon quasiparticles, free phonon quasiparticles, and magnon-polaron quasiparticles. 
Each of these quasiparticles may carry heat and spin angular momentum, manifested in thermal Hall and spin Nernst effects, respectively.

In our 2D honeycomb lattice model, the noninteracting phonon spectrum (see Fig. \ref{fig: phonon_bands}) hosts four types of gapless Dirac-like points \cite{TopologicalPhonons,PhysRevLett.131.116602}. These crossing points are sources of large Berry curvatures but since they are degenerate points thanks to the time-reversal symmetry, they do not contribute to the intrinsic thermal Hall and spin Nernst effects \cite{PhysRevB.104.045139}. On the other hand, in our magnetic model, the noninteracting magnon spectrum, see Fig \ref{fig: magnon_bands}, hosts topological gaps with nonzero Chern numbers at the K points as long as there is a finite projection of the ground-state magnetization vector onto the OOP nnn DM vector, that do contribute to the total intrinsic thermal Hall and spin Nernst effects.

Therefore, in this paper, we only focus on the intrinsic contribution to the anomalous thermal Hall and spin Nernst effects, emerging from the magnon-polaron quasiparticles.

\subsection{Anomalous thermal Hall effect}
\begin{figure}
    \centering
    \includegraphics[width=1\linewidth]{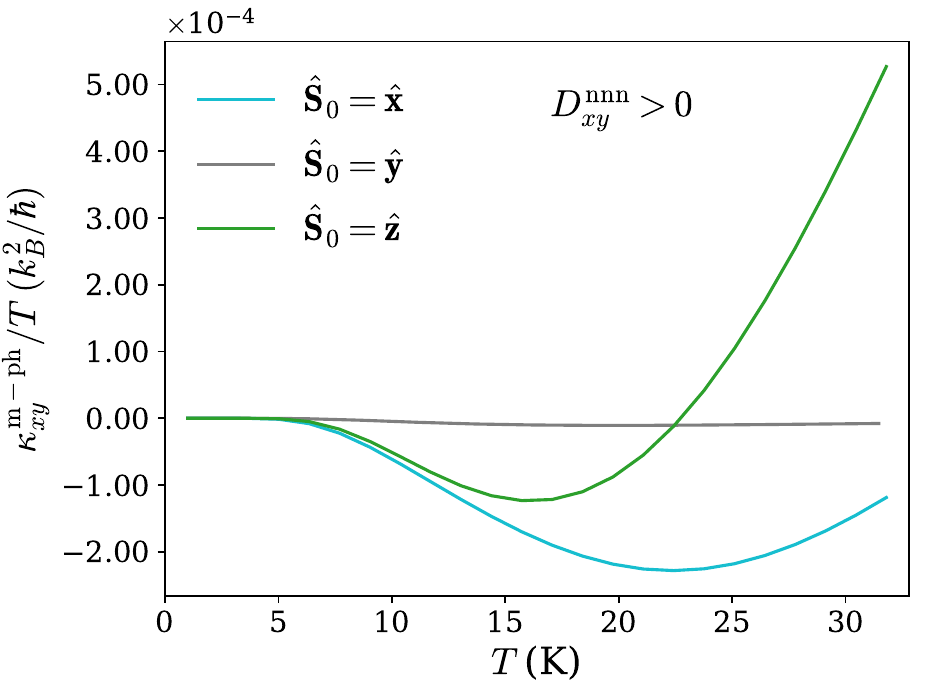}
    \caption{The anomalous thermal Hall conductivity of magnon-polaron quasiparticles, generated by a finite IP nnn DM, $D_{xy}^{\text{nnn}} = 0.3$ meV, as a function of temperature for different ground-state magnetization directions. The parameters are the same as in Fig. \ref{fig: magnon_polaron_bands}-b.}
    \label{fig: THC_Dnnn}
\end{figure}
 \begin{figure}
    \centering
        \includegraphics[width=1\linewidth]{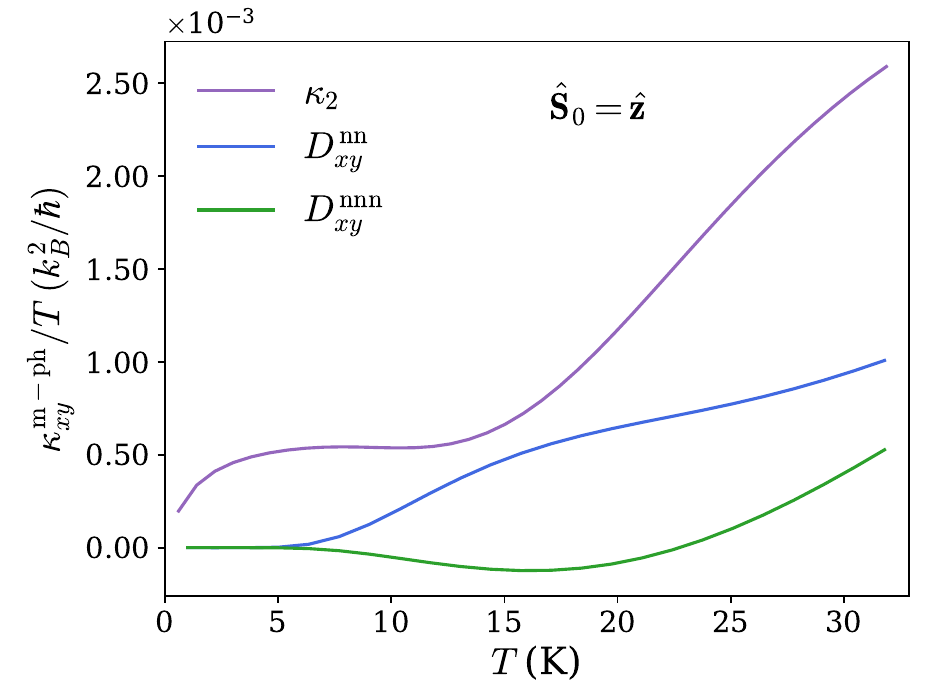}
    \caption{The anomalous thermal Hall conductivity of magnon-polaron quasiparticles as a function of temperature for different m-ph coupling mechanisms. The scenario considers an OOP ground-state magnetization. We set $\kappa_2=1.04$ meV, $D_{xy}^{\text{nn}}=0.173$ meV, and $D_{xy}^{\text{nnn}}=0.3$ meV.}
    \label{fig: THC_z}
\end{figure}
The 2D anomalous thermal Hall conductivity $\kappa_{xy}$ is a quantity that relates the transverse heat current to the applied temperature gradient, ${J}_x^Q =- \kappa_{xy} {{\nabla}}_y T$. 
Within the linear response theory, the anomalous thermal Hall conductivity is related to the Berry curvature as \cite{2023arXiv230504830Z,doi:10.7566/JPSJ.86.011010,PhysRevB.89.054420,Zhang_2016},
\begin{align}
\label{eqn: thermall_Hall_conductivity}
    \kappa_{xy} = - \frac{k_B^2 T}{\hbar \mathcal{A}} \sum_{\bm{k}} \sum_{n=1}^{N_d} c_2(g_{k,n}) \Omega_n(\bm{k}),
\end{align}
where $k_B$ is the Boltzmann constant, $T$ is the temperature, $\mathcal{A}$ is the area of the system, 
$g_{k,n}=\big(e^{\mathcal{E}_{k,n}/k_B T} - 1\big)^{-1}$ is the equilibrium Bose-Einstein distribution, and $c_2(x)= (1+x) \left(\ln{\frac{1+x}{x}}\right)^2 - \left(\ln{x}\right)^2 - 2 \text{Li}_2 (-x)$, where Li$_2(x)$ is the polylogarithm function of second order. 

As we mentioned earlier, in the absence of m-ph couplings, the only source of thermal Hall conductivity in our model is a finite $D_{z}^{\text{nnn}}$. This noninteracting conductivity has a pure magnonic origin and is finite only if the magnetization has a finite projection along the OOP nnn DM vector, i.e., the $z$ direction. However, we have shown that m-ph interactions may generate a finite Berry curvature in different phonon-like and magnon-like bands of the magnon-polaron hybrid states. Therefore, we expect a finite thermal Hall conductivity response, mediated by magnon-polaron quasiparticles, in addition.

Figure \ref{fig: THC_Dnnn} presents the thermal Hall conductivity arising from the m-ph coupling via $D_{xy}^{\text{nnn}}$, as a function of temperature for several magnetization directions. This figure shows that the thermal Hall conductivity is tunable by applying a magnetic field along different directions. The thermal Hall conductivity exhibits a varying sign for both $x$ and $z$ directions.
In the low temperature regime, the interaction between the LA phonon band and the lower magnon band $E_-$ takes precedence. However, with increasing temperature, additional bands are thermally activated, resulting in a significant alteration in their signatures.

In contrast, the anomalous thermal Hall conductivity vanishes with magnetization along the $y$ direction. This arises from the complete balance in the distribution of the Berry curvature, resulting in topologically trivial bands. Specifically, for each band gap that contributes to the conductivity, there exists a corresponding band gap elsewhere in the Brillouin zone, where the bands exhibit opposite Berry curvature, leading to a negative contribution. Consequently, the net conductivity should be zero. This finding strengthens the idea that nonzero thermal Hall conductivity can serve as an indicator of a nontrivial topology in FM systems.
It is worth mentioning that using magnetic point group analysis, we can also explain the absence of a thermal Hall response in this case. It has been demonstrated that the magnetization vector and the components of the transverse heat conductivity transform in a similar manner under magnetic point group symmetry operations \cite{PhysRevB.99.014427,PhysRevLett.126.147201}. 
If the magnetization is oriented along the $y$ direction in our model, the magnetic point group is reduced to 2/m, including a twofold rotation axis (2) and a mirror plane (/m). The combined time-reversal and spin-rotation symmetry is maintained, resulting in the Berry curvature in the momentum space being antisymmetric under this symmetry. As a consequence, both the Chern number and the thermal Hall effect are zero in this case \cite{PhysRevB.99.014427}.
In general, considering $\phi$ as the azimuthal angle of the magnetization direction in the laboratory coordinate frame, see Fig. \ref{fig: honeycomb_figure}, it follows that $\kappa_{xy}(\phi) = \kappa_{xy}(2\pi/3-\phi)$ and $\kappa_{xy}(\phi) = - \kappa_{xy}(\phi+\pi/3)$. Therefore, we get $\kappa_{xy}\big(\phi=(2k-1)\pi/6\big)=0$, with $k=1,2,3, ...$, in agreement with 2/m symmetry.

In Fig. \ref{fig: THC_z}, we compare the thermal Hall conductivities of magnon-polaron states arising from different m-ph coupling mechanisms. Unlike the DM-induced m-ph coupling, the anisotropy contribution exhibits nonzero values even at very low temperatures. One possible explanation for this disparity is the emergence of a slightly negative Berry curvature in the ZA phonon branch around the $\Gamma$-points in the presence of finite $\kappa_2$, leading to a positive conductivity. 
Interestingly, this effect diminishes when the quadratic low-energy dispersion of the ZA phonon mode is replaced by a linear dispersion, as commonly found in 3D materials or 2D honeycomb lattices with broken sublattice (rotation) symmetry. In such cases, for m-ph coupling arising from $\kappa_2$, a vanishing conductivity is found at low temperatures. 

\subsection{Spin Nernst effect}
\begin{figure}
    \centering
        \centering
        \includegraphics[width=1\linewidth]{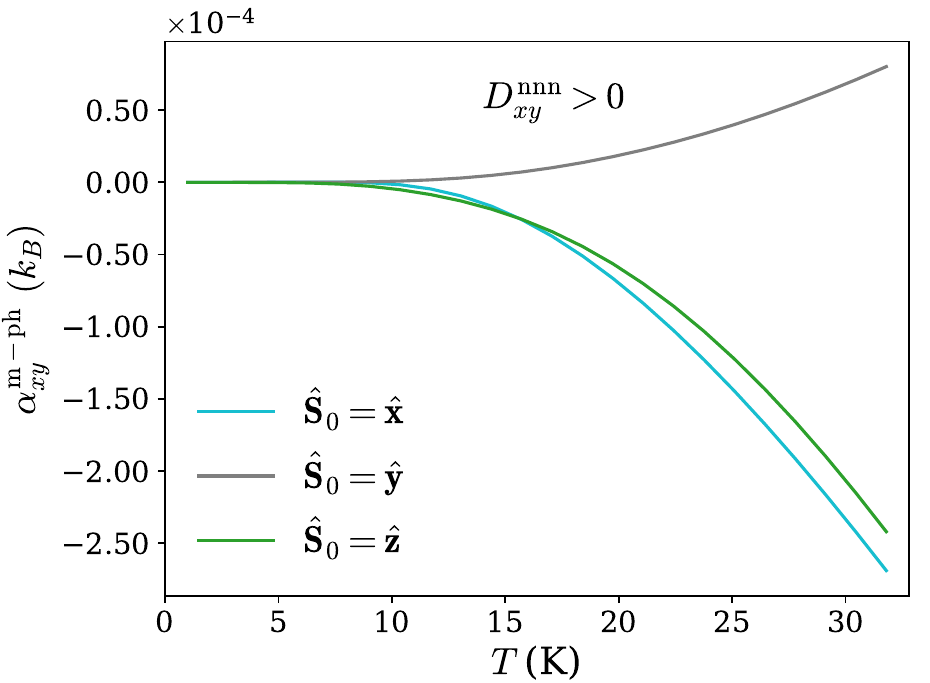}
        \label{fig: SNE_Dnnn}
    \caption{The spin Nernst coefficient of magnon-polaron quasiparticles as a function of temperature for different ground-state magnetization directions with $D_{xy}^{\text{nnn}} = 0.3$ meV. }
    \label{fig: SNE} 
\end{figure}
\begin{figure}
    \centering
        \centering
        \includegraphics[width=1\linewidth]{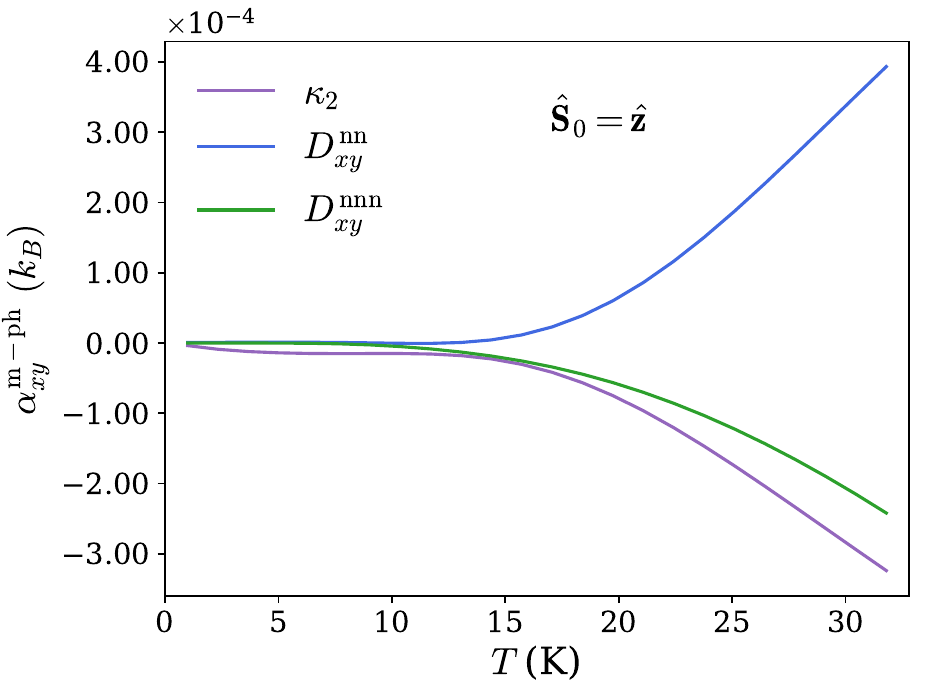}
    \caption{The spin Nernst coefficient of magnon-polaron quasiparticles as a function of temperature for different m-ph coupling mechanisms. The ground-state magnetization is OOP.  We set $\kappa_2 =  1.04$meV, $D_{xy}^{\text{nn}} = 0.173$ meV, and $D_{xy}^{\text{nnn}} = 0.3$ meV.}
    \label{fig: SNE_z}
\end{figure}
The spin Nernst coefficient $\alpha_{xy}$ relates the spin current density to the applied temperature gradient, ${J}_x^s =- \alpha_{xy} {\nabla}_y T$. Within the linear response theory, the spin Nernst coefficient is related to the \emph{spin} Berry curvature and is given by \cite{Thermal_Hall_Effect_Spin_Nernst_Effect_Spin_Density,PhysRevB.106.125103,PhysRevResearch.2.013079},
\begin{align}
\label{eqn: spin_Nernst_coefficient}
    \alpha_{xy} = -\frac{2 k_B}{\mathcal{A}} \sum_{\bm{k}} \sum_{n=1}^{N_d} c_1(g_{k,n}) \Omega_n^s(\bm{k}) ,
\end{align}
where the spin Berry curvature is, 
\begin{align}
    \Omega_n^s (\bm{k}) = 2 i \hbar^2 \sum_{\substack{m= 1 \\ m\neq n}}^{2N_d} (\sigma_3)_{nn} (\sigma_3)_{mm} \frac{\langle n_{\bm{k}} |j^s_x|{m_{\bm{k}}}\rangle \langle{m_{\bm{k}}}|v_y|{n_{\bm{k}}}\rangle}{(\Bar{\mathcal{E}}_{k,n} - \mathcal{\Bar{E}}_{k,m})^2},
\end{align}
with $ c_1(x) = (1+x)\ln{(1+x)} - x \ln{(x)}$, and  $\bm{j}^s = \frac{1}{4} \{\bm{v} , \sigma_3\mathcal{S} \}$ is the spin current operator. Moreover, $\mathcal{S}$ is the spin excitation operator, and can explicitly be written in the form $\mathcal{S} = \text{diag}(S_1, \dots, S_{N_d}) \otimes I_{2 \times 2}$, where $S_n$ is the expectation value of the spin angular momentum in the noninteracting band $n$ \cite{PhysRevB.106.125103,PhysRevResearch.2.013079}.

In Fig. \ref{fig: SNE}, we compare the spin Nernst coefficient for various magnetization directions as a function of temperature when the m-ph coupling mechanism is through $D_{xy}^{\text{nnn}}$. 
The magnitude of the coefficient is of the same order in all cases, but the sign varies. A positive coefficient is observed when the magnetization is oriented along the $y$ direction, while negative contributions are observed for the $x$ and $z$ directions. In contrast to the thermal Hall conductivity, shown in Fig. \ref{fig: THC_Dnnn}, the spin Nernst coefficient does not change sign when increasing the temperature. This difference can be attributed to the distinct distribution of the Berry curvature and the \emph{spin} Berry curvature.

Notably, even in the case where the bands are topologically trivial and the anomalous thermal Hall effect vanishes, i.e., the ground-state magnetization is along the $y$ direction, we still find a nonzero spin Nernst coefficient. The nonvanishing spin current arises from the \emph{spin} Berry curvature, which, unlike the Berry curvature, is not directly linked to the Chern number. Hence, the FM system exhibits a spin Nernst current, while the thermal Hall current is absent. Having a finite transverse spin-polarized current with a polarization along the $z$ direction, is also in agreement with the 2/m magnetic point group analysis for spin conductivity tensors \cite{PhysRevB.92.155138}.
This means that, in this case, we only have a transverse net spin current and no heat current.

In Fig. \ref{fig: SNE_z}, we compare the spin Nernst conductivity of magnon-polaron states arising from different types of m-ph couplings.
This figure illustrates that distinct sign and temperature-dependent behavior of the spin Nernst signal can be used to distinguish between different m-ph coupling mechanisms in experiments.

\section{Summery and concluding remarks}\label{summary}
We examine the effects of an external magnetic field and various m-ph coupling mechanisms on emerging magnon-polaron hybrid states. We have shown that the (spin) Berry curvature and the topology of the hybrid bands can be tuned by the direction of the applied magnetic field. We also explored the impact of the (spin) Berry curvature on thermal Hall and spin Nernst effects. We showed that the thermal Hall response of magnon-polaron hybrid states can be eliminated, while the spin Nernst effect remains finite in the system. 
Our study suggests that measuring the magnetic field dependence of anomalous Hall effects can be used as a probe of the underlying m-ph coupling and the topology of the system.
Furthermore, we suggest that, in order to discriminate between quantum transport contributions associated with magnon-polaron and free magnon quasiparticles, an angular-dependent analysis of the magnetic field is essential in the experimental setup.

\section*{Acknowledgment}
We thank Se Kwon Kim, Verena Brehm, and Jiangxu Li for helpful discussions. We also thank the anonymous referees for their valuable comments.
This project has been supported by Norwegian Financial Mechanism Project No. 2019/34/H/ST3/00515, ``2Dtronics''; and partially by the Research Council of Norway through its Centers of Excellence funding scheme, Project No. 262633, ``QuSpin''.

\bibliography{Refs}

\onecolumngrid
\appendix
\section{Rotation matrix} \label{AppA}
To find the magnon dispersion for an arbitrary direction of the spins, it is convenient to change the frame of reference from the lab reference to $\{\hat{\bm{e}}_1,\hat{\bm{e}}_2, \hat{\bm{e}}_3\}$, where $\hat{\bm{e}}_i$ is the unit vector. The new frame of reference is defined such that $\hat{\bm{e}}_3$ is aligned with the magnetic field, and we assume a strong magnetic field such that the ground state spin direction also aligns along $\hat{\bm{e}}_3$. This frame of reference is related to the lab-frame via a rotation in the form \cite{PhysRevX.11.031047}
\begin{equation}
    \hat{\bm{e}}_1 = \mathcal{R} \hat{\bm{x}}, \hspace{2mm}
    \hat{\bm{e}}_2 = \mathcal{R} \hat{\bm{y}}, \hspace{2mm}
    \hat{\bm{e}}_3 = \mathcal{R} \hat{\bm{z}}, \\
\end{equation}
where $\mathcal{R} = \mathcal{R}_z(\phi) \mathcal{R}_y(\theta) \mathcal{R}_x(\vartheta)$ is the rotation matrix and
\begin{gather}
    \mathcal{R}_z(\phi) = \begin{bmatrix}
        \cos(\phi) & -\sin(\phi) & 0 \\
        \sin(\phi) &\cos(\phi) & 0 \\
        0 & 0 & 1
    \end{bmatrix}, \hspace{2mm} 
    \mathcal{R}_y(\theta) = \begin{bmatrix}
        \cos(\theta) & 0 & \sin(\theta)\\
        0 & 1 &0 \\
        -\sin(\theta) & 0 &\cos(\theta)
    \end{bmatrix}, \hspace{2mm}
    \mathcal{R}_x(\vartheta) = \begin{bmatrix}
        1 & 0 & 0 \\
        0 & \cos(\vartheta) & -\sin(\vartheta)\\
        0 & \sin(\vartheta) & \cos(\vartheta)
    \end{bmatrix}. 
\end{gather}
Denoting $\bm{S}'_i$ as the spins in the new frame of reference, they are related to the ones in the lab-frame by $\bm{S}'_i = \mathcal{R}^T \bm{S}_i$, where
\begin{gather}
    \mathcal{R} = \begin{bmatrix}
        \mathcal{R}_{x}^1 & \mathcal{R}_{x}^2 & \mathcal{R}_{x}^3 \\
        \mathcal{R}_{y}^1 & \mathcal{R}_{y}^2 & \mathcal{R}_{y}^3 \\
        \mathcal{R}_{z}^1 & \mathcal{R}_{z}^2 & \mathcal{R}_{z}^3 \\
    \end{bmatrix}  
    =  
    \begin{bmatrix}
    \cos(\theta)\cos(\phi) & \sin(\theta)\cos(\phi)\sin(\vartheta)-\sin(\phi)\cos(\vartheta) & \sin(\theta)\cos(\phi)\cos(\vartheta) + \sin(\phi)\sin(\vartheta) \\
    \cos(\theta)\sin(\phi) & \sin(\theta)\sin(\phi)\sin(\vartheta) + \cos(\phi)\cos(\vartheta) & \sin(\theta)\sin(\phi)\cos(\vartheta) - \cos(\phi)\sin(\vartheta) \\ 
    -\sin(\theta) & \cos(\theta)\sin(\vartheta) & \cos(\theta)\cos(\vartheta) \\
    \end{bmatrix}.        
\end{gather}
The rotation matrix is an orthogonal matrix satisfying $\mathcal{R}^T \mathcal{R}=I$. For the spins in the new frame of reference, the Holstein-Primakoff transformation for large spin limit can then be written as
\begin{equation}
        S_{iA1}' \approx \sqrt{\frac{S}{2}} (a_i + a_i^\dagger), \hspace{2mm} 
        S_{iA2}' \approx - i \sqrt{\frac{S}{2}} (a_i - a_i^\dagger), \hspace{2mm}
        S_{iA3}' = S - a_i^\dagger a_i,
\end{equation}
and similarly for the spins at sublattice $B$.

\section{Magnon-phonon interactions}\label{AppB}
In this appendix, we derive the effective m-ph interaction Hamiltonian for various m-ph coupling mechanisms.

\subsection{Contribution from the NNN DM interaction}\label{AppB-1}
The nnn DM vector is expressed in the following form:
\begin{equation}
    \bm{D}^{\mathrm{nnn}} (\bm{r}_{ij}) = -\eta_{ij} D_{xy}^{\mathrm{nnn}}(\bm{r}_{ij}) \hat{\bm{r}}_{ij} + \nu_{ij} D_z^{\mathrm{nnn}}(\bm{r}_{ij}) \hat{\bm{z}}.
\end{equation}
We defined $\bm{r}_{ij} = \bm{r}_i - \bm{r}_j$, where $\bm{r}_i$ denotes the instantaneous position vector at lattice site $i$, defined as $\bm{r}_i = \bm{R}_i + \bm{u}_i$ with $\bm{R}_i$ as the equilibrium position and $\bm{u}_i$ as the displacement. By Taylor expanding the interaction strengths $D_{xy}^{\mathrm{nnn}}(\bm{r}_{ij})$, $D_z^{\mathrm{nnn}}(\bm{r}_{ij})$ and the unit vector $\hat{\bm{r}}_{ij}$ around the equilibrium position $\bm{R}_{ij} = \bm{R}_i - \bm{R}_j$,
\begin{equation}
    \begin{split}
        & D_{xy}^{\text{nnn}} (\bm{r}_{ij}) \approx D_{xy}^{\text{nnn}} (\bm{R}_{ij}) + \left. \nabla_{\bm{r}_{ij}} D_{xy}^{\text{nnn}}  (\bm{r}_{ij})\right|_{\boldsymbol{R}_{ij}}  \cdot \bm{u}_{ij}, \\
        & D_{z}^{\text{nnn}} (\bm{r}_{ij}) \approx D_{z}^{\text{nnn}} (\bm{R}_{ij}) + \left. \nabla_{\bm{r}_{ij}} D_{z}^{\text{nnn}}  (\bm{r}_{ij})\right|_{\boldsymbol{R}_{ij}}  \cdot \bm{u}_{ij}, \\
        & \hat{\bm{r}}_{ij} \approx \frac{\bm{R}_{ij}+ \bm{u}_{ij}}{|\bm{R}_{ij} + \bm{u}_{ij}|} = \frac{\bm{R}_{ij} + \bm{u}_{ij}}{\sqrt{\bm{R}_{ij}^2 + 2\bm{R}_{ij}\cdot \bm{u}_{ij} + \bm{u}_{ij}^2}} \approx \frac{\bm{R}_{ij}+\bm{u}_{ij}}{|\bm{R}_{ij}|}(1 - \frac{\bm{R}_{ij}\cdot \bm{u}_{ij}}{|\bm{R}_{ij}|^2}),
    \end{split}
\end{equation}
the nnn DM vector can be expressed as a first-order approximation in terms of the ionic displacement as
\begin{equation}
    \begin{split}
        \bm{D}^{\mathrm{nnn}}(\bm{r}_{ij}) \approx& \bm{D}^{\mathrm{nnn}}(\bm{R}_{ij}) - \eta_{ij} \frac{D_{xy}^{\mathrm{nnn}}(\bm{R}_{ij})}{|\bm{R}_{ij}|} \left(\bm{u}_{ij} - (\hat{\bm{R}}_{ij} \cdot \bm{u}_{ij}) \hat{\bm{R}}_{ij} \right) \\
        & - \eta_{ij} \left( \left. \nabla_{\boldsymbol{r}_{ij}} D_{xy}^{\mathrm{nnn}}(\bm{r}_{ij}) \right|_{\boldsymbol{R}_{ij}} \cdot \bm{u}_{ij} \right) \hat{\bm{R}}_{ij} + 
        \nu_{ij} \left( \left. \nabla_{\boldsymbol{r}_{ij}} D_z^{\mathrm{nnn}}(\bm{r}_{ij}) \right|_{\boldsymbol{R}_{ij}} \cdot \bm{u}_{ij} \right) \hat{\bm{z}} \\
        \approx& \bm{D}^{\mathrm{nnn}}(\bm{R}_{ij}) - \eta_{ij} \frac{D_{xy}^{\mathrm{nnn}}(\bm{R}_{ij})}{|\bm{R}_{ij}|} \left(\bm{u}_{ij} - (\hat{\bm{R}}_{ij} \cdot \bm{u}_{ij}) \hat{\bm{R}}_{ij} \right),
    \end{split}
\end{equation}
where we also defined $\bm{u}_{ij} = \bm{u}_i - \bm{u}_j$. In the last line, we only keep the lowest order term and ignored the gradient terms. This term makes a contribution to the m-ph interacting Hamiltonian, 
\begin{equation}
    \begin{split}
\mathcal{H}_{\mathrm{D}^{\mathrm{nnn}}_{xy}} =& \sum_{\langle \langle i,j \rangle \rangle} \sum_{\mu,\nu} (u_{i \mu} - u_{j\mu}) \eta_{ij} \frac{D_{xy}^{\mathrm{nnn}}}{|\bm{R}_{ij}|} \left(\delta_{\mu\nu}- \hat{R}_{ij}^{\mu} \hat{R}_{ij}^{\nu} \right) (\bm{S}_i \times \bm{S}_j)_{\nu},
    \end{split}
\end{equation}
with $D_{xy}^{\mathrm{nnn}} = D_{xy}^{\mathrm{nnn}}(\boldsymbol{R}_{ij})$. In terms of the new frame of reference, $(\bm{S}_i \times \bm{S}_j)_{\nu} = (\mathcal{R} (\bm{S}'_i \times \bm{S}'_j))_{\nu}$, the Hamiltonian can be expressed in a concise form as follows
\begin{equation}
\mathcal{H}_{\mathrm{D}^{\mathrm{nnn}}_{xy}} = \sum_{\langle \langle i,j \rangle \rangle} \sum_{\mu,\nu} (u_{i\mu} - u_{j\mu}) F_{ij}^{\mu\nu} (\bm{S}'_i \times \bm{S}'_j)_{\nu},
\end{equation}
where $\mu \in \{x,y,x\}$, $\nu \in \{1,2,3\}$ and the coupling matrix is defined by
\begin{equation}
    F_{ij}^{\mu\nu} = \sum_{\xi=x,y,z} \eta_{ij} \frac{D_{xy}^{\mathrm{nnn}}}{|\bm{R}_{ij}|} \left(\delta_{\mu\xi}- \hat{R}_{ij}^{\mu} \hat{R}_{ij}^{\xi} \right) \mathcal{R}_{\xi}^{\nu}.
\end{equation}
Writing out the summation, the coupling to IP ($\mu = x,y$) and OOP ($\mu = z$) phonon modes can be separated as
\begin{equation}
    \begin{split}
        (\mu = x,y): \hspace{2mm} F_{ij}^{\mu\nu} =& \eta_{ij} \frac{D_{xy}^{\mathrm{nnn}}}{|\bm{R}_{ij}|} \left[ (\delta_{\mu x} - \hat{R}_{ij}^{\mu} \hat{R}_{ij}^{x} ) \mathcal{R}_{x}^{\nu} + (\delta_{\mu y} - \hat{R}_{ij}^{\mu} \hat{R}_{ij}^{y}) \mathcal{R}_{y}^{\nu} ) \right], \\ 
        (\mu = z): \hspace{2mm} F_{ij}^{z\nu} =& \eta_{ij} \frac{D_{xy}^{\mathrm{nnn}}}{|\bm{R}_{ij}| }  \mathcal{R}_{z}^{\nu}.
    \end{split}
\end{equation}

\subsection{Contribution from the NN DM interaction}\label{AppB-2}
The nn DM vector is in the form
\begin{equation}
    \bm{D}^{\mathrm{nn}} (\bm{r}_{ij}) = D_{xy}^{\mathrm{nn}} (\bm{r}_{ij}) (\hat{\bm{z}} \times \hat{\bm{r}}_{ij}) + D_z^{\mathrm{nn}} (\bm{r}_{ij}) \hat{\bm{z}}.
\end{equation}
By performing a Taylor expansion, the DM vector's lowest order contributions are
\begin{equation}
    \begin{split}
        \bm{D}^{\mathrm{nn}}(\bm{r}_{ij}) \approx & \bm{D}^{\mathrm{nn}}(\bm{R}_{ij}) + \frac{D_{xy}^{\mathrm{nn}}}{|\bm{R}_{ij}| } \left(\hat{\bm{z}} \times (\bm{u}_{ij} - (\hat{\bm{R}}_{ij} \cdot \bm{u}_{ij}) \hat{\bm{R}}_{ij}) \right) \\
        & + \left( \left. \nabla_{\boldsymbol{r}_{ij}} D_{xy}^{\mathrm{nn}} (\bm{r}_{ij}) \right|_{\boldsymbol{R}_{ij}} \cdot \bm{u}_{ij} \right) (\hat{\bm{z}} \times \hat{\bm{R}}_{ij}) + \left( \left. \nabla_{\boldsymbol{r}_{ij}} D_{z}^{\mathrm{nn}} (\bm{r}_{ij}) \right|_{\boldsymbol{R}_{ij}} \cdot \bm{u}_{ij} \right) \hat{\bm{z}} \\
        \approx & \bm{D}^{\mathrm{nn}}(\bm{R}_{ij}) + \frac{D_{xy}^{\mathrm{nn}}}{|\bm{R}_{ij}| } \left(\hat{\bm{z}} \times (\bm{u}_{ij} - (\hat{\bm{R}}_{ij} \cdot \bm{u}_{ij}) \hat{\bm{R}}_{ij}) \right),
    \end{split}
\end{equation}
with $D_{xy}^{\mathrm{nn}} = D_{xy}^{\mathrm{nn}}(\bm{R}_{ij})$ and we again ignore the terms containing derivatives in the last line. Therefore the m-ph interaction reads,
\begin{equation}
\mathcal{H}_{\mathrm{D}^{\mathrm{nn}}_{xy}} = \sum_{\langle i,j \rangle} \sum_{\mu, \nu} (u_{i\mu}- u_{j\mu}) T_{ij}^{\mu\nu} (\bm{S}'_i \times \bm{S}'_j)_{\nu},
\end{equation}
for $\mu \in \{x,y,x\}$ and $\nu \in \{1,2,3\}$, and the nearest-neighbor coupling matrix is defined by
\begin{equation}
    T_{ij}^{\mu\nu} = -\sum_{\substack{\xi,\gamma \\ \in \{x,y,z\}}} \frac{D_{xy}^{\mathrm{nn}}}{|\bm{R}_{ij}| } \varepsilon_{z\gamma \xi} \left(\delta_{\mu\gamma} - \hat{R}_{ij}^{\mu} \hat{R}_{ij}^{\gamma} \right) \mathcal{R}_{\xi}^{\nu},
\end{equation}
where $\varepsilon_{z\gamma\xi}$ is the Levi-Civita tensor. The lowest order contribution does not result in coupling with OOP phonon modes. However, the coupling with IP phonon modes takes the following form
\begin{equation}
    \begin{split}
        (\mu = x,y):  \hspace{2mm} T_{ij}^{\mu\nu} =& -\frac{D_{xy}^{\mathrm{nn}}}{|\bm{R}_{ij}| } \left[ (\delta_{\mu x} - \hat{R}_{ij}^{\mu} \hat{R}_{ij}^{x}) \mathcal{R}_{y}^{\nu} - (\delta_{\mu y} - \hat{R}_{ij}^{\mu} \hat{R}_{ij}^{y}) \mathcal{R}_{x}^{\nu} \right].
    \end{split}
\end{equation}

\subsection{Contribution from magnetocrystalline anisotropy}\label{AppB-3}
The magnetoelastic energy caused by crystalline anisotropy occurs because the movement of atoms results in a local alteration of crystal axes, which in turn affects the crystalline anisotropy and becomes linked to spins. As crystalline anisotropy is a feature of all solids, this form of magnetoelastic coupling should also exist. The derivation of the contribution to the interacting Hamiltonian takes its starting point in the anisotropy energy density in an untrained cubic crystal. Expanding the energy density in terms of the strains, the magnetoelastic energy density is in the form \cite{RevModPhys.21.541, PhysRevLett.123.237207,PhysRevB.89.184413}
\begin{equation}
    f^{\mathrm{me}} = b_1 (\hat{m}_x^2 e_{xx} + \hat{m}_y^2 e_{yy} + \hat{m}_z^2 e_{zz}) + 2b_2 (\hat{m}_x \hat{m}_y e_{xy} + \hat{m}_{y} \hat{m}_z e_{yz} + \hat{m}_x \hat{m}_z e_{zx}),
\end{equation}
where $e_{\mu\nu} = \frac{1}{2} \left( \partial_{\nu} u_{\mu} + \partial_{\mu} u_{\nu} \right)$ are the strains and $\partial_{\nu} u_{\mu} = \frac{\partial u_{\mu}}{\partial r_{\nu}}$. The magnetoelastic coupling constants are denoted by $b_1$ and $b_2$, while the directional cosines of the magnetization are represented by $(\hat{m}_{x}, \hat{m}_{y}, \hat{m}_{z})$, with $\hat{m}_\mu=S_{i\mu}/S$. Additionally, in the continuum limit, $\bm{u}(\bm{r})$ signifies the displacement. Since we look at a two-dimensional system, we can neglect the derivative of the OOP displacements \cite{PhysRevLett.123.237207}. By rotating the Cartesian magnetization components to the new frame of reference and neglecting higher order contributions, we obtain
\begin{equation}
    \begin{split}
        f^{\mathrm{me}} \approx & 2b_1 \left[ \partial_x u_x (\Gamma^1_{xx} \hat{m}'_1 + \Gamma^2_{xx} \hat{m}'_2) + \partial_y u_y ( \Gamma^1_{yy} \hat{m}'_1 + \Gamma^2_{yy} \hat{m}'_2) \right] \\
        & +2 b_2 \left[(\partial_y u_x + \partial_x u_y)(\Gamma^1_{xy} \hat{m}'_1 + \Gamma^2_{xy} \hat{m}'_2) + \partial_x u_z (\Gamma^1_{xz} \hat{m}'_1 + \Gamma^2_{xz} \hat{m}'_2)  \right. 
         \left. + \partial_y u_z (\Gamma^1_{yz} \hat{m}'_1 + \Gamma^2_{yz} \hat{m}'_2) \right],
    \end{split}
\end{equation}
where $\Gamma^{\nu}_{\mu\mu'} =  (\mathcal{R}_{\mu}^{\nu} \mathcal{R}_{\mu'}^3 + \mathcal{R}_{\mu}^3 \mathcal{R}_{\mu'}^{\nu})/2$ and we utilized $\hat{m}'_3 \approx 1$. In order to use this expression in a discrete lattice, we approximate the strain tensor with the discrete strain tensor,
\begin{equation}
    \Tilde{e}_{\mu\nu} = \frac{1}{2} \frac{1}{|\bm{R}_i - \bm{R}_j|^2}\left[(R_{i\nu} - R_{j\nu}) (u_{i\mu} - u_{j\mu}) + (R_{i\mu} - R_{j\mu}) (u_{i\nu} - u_{j\nu}) \right],
\end{equation}
which is proportional to the strain tensor in the long wavelength limit. By using $\hat{m}'_{\mu} = S'_{i\mu}/S$ and summing over the entire lattice as well as the nearest neighbors \cite{PhysRevLett.123.237207},
\begin{equation}
    F_{\text{me}} = \sum_{i  } f^{\text{me}}_{i} a^3,
\end{equation}
we obtain the magnetoelastic energy. Thus the contribution to the Hamiltonian can be written as

\begin{equation}
    \begin{split}
        \mathcal{H}_{\mathrm{me}} &= \sum_{\langle i,j \rangle} \frac{1}{|\bm{R}_{ij}|^2} \left[ \kappa_1 \left(R_{ij}^x u_{ij}^x (\Gamma^1_{xx} S'_{i1} + \Gamma^2_{xx} S'_{i2}) + R_{ij}^y u_{ij}^y ( \Gamma^1_{yy} S'_{i1} + \Gamma^2_{yy} S'_{i2}) \right) \right. \\
        & + \kappa_2 \left( (R_{ij}^y u_{ij}^x + R_{ij}^x u_{ij}^y) (\Gamma^1_{xy} S'_{i1} + \Gamma^{2}_{xy} S'_{i2}) + R_{ij}^x u_{ij}^z (\Gamma^1_{xz} S'_{i1} + \Gamma^2_{xz} S'_{i2}) \right) \\
        & + \quad \left. \left. R_{ij}^y u_{ij}^z (\Gamma^1_{yz} S'_{i1} + \Gamma^2_{y2} S'_{i2} \right) \right] \\
        &= \sum_{\langle i,j \rangle} \sum_{\mu\nu} (u_{i\mu} - u_{j\mu}) K_{ij}^{\mu\nu} S'_{i\nu},
    \end{split}
\end{equation}
for $\mu \in \{x,y,z\}$ and $\nu \in \{1,2\}$. The interaction strengths $\kappa_1 = 2 b_1 a^3/S$ and $\kappa_2 = 2 b_2 a^3/S$ have been defined, each expressed in units of energy. The coupling matrix is defined as follows

\begin{equation}
    K_{ij} = \frac{1}{|\bm{R}_{ij}|^2} \begin{bmatrix}
        \kappa_1 R_{ij}^x \Gamma^1_{xx} + \kappa_2 R_{ij}^y \Gamma^1_{xy} & \kappa_1 R_{ij}^x \Gamma^2_{xx} + \kappa_2 R_{ij}^y \Gamma^2_{xy} \\
        \kappa_1 R_{ij}^y \Gamma^1_{yy} + \kappa_2 R_{ij}^x \Gamma^1_{xy} & \kappa_1 R_{ij}^y \Gamma^2_{yy} + \kappa_2 R_{ij}^x \Gamma^2_{xy} \\
        \kappa_2 (R_{ij}^x \Gamma^1_{xz} + R_{ij}^y \Gamma^1_{yz}) & \kappa_2( R_{ij}^x \Gamma^2_{xz} + R_{ij}^y \Gamma^2_{yz})
    \end{bmatrix}.
\end{equation}

When the magnetization is OOP, the expression simplifies to
\begin{equation}
    (\mathrm{magnetization}\perp \mathrm{plane}): \hspace{3mm} \mathcal{H}_{\mathrm{me}} = \sum_{\langle i,j \rangle} \frac{\kappa_2}{2 |\bm{R}_{ij}|} (u_{iz} - u_{jz}) (\hat{\bm{R}}_{ij} \cdot \bm{S}_i).
\end{equation}
In contrast to the DM interactions, which interact solely with IP phonon modes when the magnetization is OOP, the anisotropy induces a coupling that is limited to OOP phonon modes for the same magnetization direction. Similarly, a magnetization aligned with the $\hat{\bm{x}}$ direction generates the following interacting Hamiltonian:
\begin{equation}
    (\mathrm{magnetization} \hspace{1mm} || \hspace{1mm} \hat{\bm{x}}): \hspace{3mm} \mathcal{H}_{\mathrm{me}} = \sum_{\langle i,j \rangle} \frac{\kappa_2}{2 |\bm{R}_{ij}|} \left((u_{ij}^x \hat{R}^y_{ij} + u_{ij}^y \hat{R}_{ij}^x) S_{iy} + u_{ij}^z \hat{R}_{ij}^x S_{iz} \right).
\end{equation}
It is worth noting that for $\kappa_1$ to have an impact on the magnon-polaron dispersion, the orientation of the magnetization must not be parallel to any of the Cartesian axes in the coordinate system. This is connected to the fact that the derivation begins with the anisotropy energy density specific to a cubic lattice.

\subsection{The effective m-ph Hamiltonian}
The total m-ph Hamiltonian that leads to a coherent coupling of magnons and phonons reads,
\begin{gather}    
    \mathcal{H}_{\mathrm{m-ph}} = \overbrace{\sum_{\langle i,j \rangle} \sum_{\mu,\nu} (u_{i\mu} - u_{j\mu}) T_{ij}^{\mu\nu} (\bm{S}'_i \times \bm{S}'_j)_{\nu}}^{\mathcal{H}_{\mathrm{D}^{\mathrm{nn}}_{xy}}} + \overbrace{\sum_{\langle \langle i,j \rangle \rangle} \sum_{\mu,\nu} (u_{i\mu} - u_{j\mu}) F_{ij}^{\mu\nu} (\bm{S}'_i \times \bm{S}'_j)_{\nu}}^{\mathcal{H}_{\mathrm{D}^{\mathrm{nnn}}_{xy}}} 
    + \overbrace{\sum_{\langle i,j \rangle} \sum_{\mu\nu} (u_{i\mu} - u_{j\mu}) K_{ij}^{\mu\nu} S'_{i\nu}}^{\mathcal{H}_{\mathrm{me}}}. 
\end{gather}
Note that by neglecting the terms that contain gradients of DM vectors, only the IP components of DM interactions, $D_{xy}^{\mathrm{nn}}$ and $D_{xy}^{\mathrm{nnn}}$, are responsible for coherent m-ph hybridization.

\end{document}